\documentclass[10pt,preprint]{aastex}

\begin{document}

\title{Relaxation of magnetic field relative to plasma density revealed from microwave zebra patterns associated with solar flares}

\author{Sijie Yu, Yihua Yan, and Baolin Tan}
\affil{ Key Laboratory of Solar Activity, National Astronomical Observatories Chinese Academy of Sciences, Beijing 100012, China.}
\email{sjyu@nao.cas.cn, yyh@nao.cas.cn, and bltan@nao.cas.cn }

\begin{abstract}
It is generally considered that the emission of microwave zebra pattern (ZP) structures requires high density and high temperatures, which is similar to the situation of the flaring region where primary energy releases. Therefore, the parameters analysis of ZPs may reveal the physical conditions of the flaring source region. This work investigates the variations of 74 microwave ZP structures observed by Chinese Solar Broadband Radio Spectrometer (SBRS/Huairou) at 2.6--3.8 GHz in 9 solar flares, and found that the ratio between the plasma density scale height $L_N$ and the magnetic field scale height $L_B$ in emission source displays a tendency of decrease during the flaring processes. The ratio $L_N/L_B$ is about 3--5 before the maximum of flares. It decreases to about 2 after the maximum. The detailed analysis of three typical X-class flares implied that the variation of $L_N/L_B$ during the flaring process is most likely due to the topological changes of the magnetic field in the flaring source region, and the step-wise decrease of $L_N/L_B$ possibly reflects the magnetic field relaxation relative to the plasma density when the flaring energy released. This result may also constrain the solar flare modeling to some extent.

\end{abstract}

\keywords{Sun: flares --- Sun: magnetic topology --- Sun: radio radiation}

Submitted to: \textit{The Astrophysical Journal} 

\section{INTRODUCTION}\label{sec-1}

It is widely accepted that the magnetic free energy stored in corona is the main energy source responsible for powering solar flares. Storage of magnetic free energy requires a non-potential magnetic field, which is always shown up as a sheared and/or twisted magnetic structure \citep{2011SSRv..158....5H}. In the processes of solar flares, magnetic free energy is converted to thermal and kinetic energy of plasma via the dissipation of electric current, accompanied with topological change of magnetic field to a lower energy state than pre-flare state \citep{2002A&ARv..10..313P,2005LRSP....2....7L, 2006ApJ...649.1064L,2011LRSP....8....6S}. This scenario can successfully interpret many observational features of solar flares. However, the detailed evolution of such coronal phenomena has not been fully understood yet, especially in the process of energy release \citep{2011LRSP....8....6S}, due to the lack of direct quantitative measurements of the coronal magnetic field in the source region \citep{2004ASSL..314...47B}.

The microwave spectral fine structures provide a unique diagnostics of the magnetic field and ambient plasma around the source. Zebra pattern (ZP) is one of the most intriguing spectral structures in the solar radio emission, which consists of a number of almost parallel and equidistant stripes superimposed on the background type IV radio burst in dynamic spectrum. They have been recorded and studied for several decades \citep{1959Natur.184..887E,1972SoPh...25..210S,1975SoPh...44..461Z,1977SvA....21..744Z} in metric \& decimetric wavelength, and later in microwave range \citep{2000A&A...364..853N,2003A&A...406.1071C,2005A&A...431.1037A,2006SSRv..127..195C,2007SoPh..241..127K,2012ApJ...744..166T}. There are several theoretical model proposed to interpret the formation of ZP structures. The model of Bernstein mode \citep[BM model:][]{1972SoPh...25..188R,1973SoPh...33..225C} is the first one to interpret such fine structure. Another important model is whistler wave model \citep[WW model:][]{1975A&A....40..405K,2006SSRv..127..195C}, based on the interaction between plasma electrostatic waves and whistler waves. The most developed model for ZP is the double plasma resonance model \citep[DPR model:][]{1975SoPh...44..461Z}. The detail information related to these theoretical models can be found in the reviews \citep{2006SSRv..127..195C,2010RAA....10..821C}. The microwave ZPs are closely related to the inhomogeneous plasma and magnetic field around flaring regions in corona by DPR model. According to this model, the emission source is considered as magnetic flux tube filled with nonequilibrium energetic particles. The anisotropic distribution of energetic particles in flux tube develops kinetic instabilities, and stimulates electrostatic waves. The excitation of dominating electrostatic upper hybrid waves are greatly enhanced at some resonance levels, where the upper hybrid frequency $f_{uh}$ is close to the harmonics of electron cyclotron frequency $f_{ce}$ in flux tube:
\begin{equation}\label{equ-0}
f_{uh}=(f_{pe}^{2}+f_{ce}^{2})^{1/2} \simeq sf_{ce}
\end{equation}
where $f_{pe}$ is the plasma frequency of electrons, and s is the harmonic number. Then these waves escape from the local regions through some nonlinear plasma processes as they are transformed to transverse waves.

It is generally considered that the emission of microwave fine structures like ZP requires high density and high temperature, which coincides with the situation of the expected primary energy release site of flare \citep{1998ARA&A..36..131B}, implies that the source of microwave ZP may be located near the flare core region where the magnetic energy is released. This also indicates that parameters analysis of ZPs may present some features of the flaring core region. \citet{2007PASJ...59S.815Y} estimated the ratio between the plasma density scale height $L_N$ and the magnetic field scale height $L_B$ derived from three ZPs in an X3.4 solar flare event on 2006 December 13 observed by the Chinese Solar Broadband Radio Spectrometer \citep[SBRS/Huairou;][]{2004SoPh..222..167F} by DPR model. It is pointed out that the $L_N/L_B$ decreases by a factor of 2 from the impulsive phase of the flare to its maximum. This is the motivation behind this work: now that the $L_N/L_B$ is closely related to the geometrical structure of coronal magnetic field, the topological change of magnetic field during energy releasing process would leave a trace on the $L_N/L_B$, i.e., ZP is an indirect tracker of the physical process of solar flare.

With the aid of high-temporal, high-spectral resolution observation data obtained at the SBRS/Huairou in 2.6--3.8 GHz \citep{2002ESASP.506..375Y,2004SoPh..222..167F}, we are encouraged to take a step forward to address this issue. In this work, we investigate the temporal variations of the $L_N/L_B$ derived from 74 ZPs appeared in the flaring processes of 9 eruptive solar flares from 1997 to 2006. This paper is organized as following. In Section 2, we describe the observations data of ZPs associated with solar flares, and present the data analysis. Section 3 presents the results. Discussions and conclusion are presented in Section 4.

\begin{deluxetable}{cccccccccccccc}
\tablecolumns{14}
\tabletypesize{\scriptsize}
\tablewidth{0pc}
\tablecaption{The list of flare events with microwave ZPs at 2.6--3.8 GHz from 1997 to 2006\label{tbl-1}}
\tablehead{
\multicolumn{4}{c}{Flare} & \colhead{} & \multicolumn{9}{c}{ZP}\\
\cline{1-4} \cline{6-14} \\
\colhead{Event}       &  \colhead{Class}          & \colhead{AR}         & \colhead{SXR peak} & &
\colhead{Start}       &   \colhead{End}           &
\colhead{f}           & \colhead{$\Delta f$}      &
\colhead{Pol}           & \colhead{No}           & \colhead{Ns}           & \colhead{Ref}           & \colhead{S}\\

\colhead{}        &   \colhead{}              & \colhead{NOAA}       & \colhead{(UT)} & &
\colhead{(UT)}        & \colhead{(UT)}            &
  \colhead{(GHz)}     & \colhead{(MHz)}           &
\colhead{}            & \colhead{}           & \colhead{}           & \colhead{}           & \colhead{}\\
}

\startdata

09.04.2000    &
M3.1                  & 8948                       & 23:42        &   &
23:35:12                   & 23:35:18           &  2.60-3.10           &    20-40                  &
STRONG R                     & 1           & 3           & 1           &$\lozenge$\\

29.10.2000           &
C4.4                  & 9209                       & 01:57        &    &
02:06:32                   & 02:35:05           &2.60-3.10             & 65-80                      &
R                     & 17           & 3-7           & 2           &$\vartriangle$     \\

24.11.2000            &
X2.0                  & 9236                       & 05:02        &    &
04:59:56                   & 05:01:57           &2.60-3.80             & 55-60                      &
L\&R                  & 4           & 3           & 2           &$\square$     \\

25.11.2000           &
M8.2                  & 9240                       & 01:31        &    &
00:59:19                   & 01:09:30           &2.60-3.80             & 50                         &
L\&R                  & 4           & 5-6           & 2           &$\blacklozenge$     \\

19.10.2001           &
X1.6                  & 9661                       & 01:05        &    &
00:51:00                   & 01:19:55           &2.60-3.00             & 55-75                      &
R                     & 18           & 3-8           & 2           &$\times$     \\

21.04.2002           &
X1.5                  & 9906                       & 01:50        &    &
01:45:40                   & 02:01:45           & 2.60-3.80            & 30-70                      &
L                     &  10           & 10-34           & 3           &$+$     \\

18.11.2003            &
M3.9                  &  10501                      & 08:31        &    &
08:22:42                   & 08:26:50           &   2.60-3.50          &   30-50                    &
STRONG R                     &  3         & 3-5           &        &$\blacktriangle$     \\

09.07.2005            &
M2.8                  &  10786                      &  22:06       &    &
22:03:16                   & 22:04:53           &  2.60-3.50           &   50-80                    &
L\&R                     &    6         & 8           &     &$\blacksquare$     \\

13.12.2006            &
X3.4                  & 10930                      & 02:40        &    &
02:22:30                   & 03:03:00           & 2.60-3.80             &  50-250                     &
R                     &   13           & 3-6           & 4           &$\ast$     \\

\enddata

\tablecomments{Start (UT)-- time of the first ZP in the event; End (UT)-- time of the last ZP in the event; AR NOAA-- number of NOAA active region; SXR peak (UT)-- time of soft X-ray peak; f (GHz)-- frequency range of ZPs; $\Delta$f (MHz)-- frequency separation of adjacent stripes; Pol-- left- or right-handed circular polarization; No-- number of ZPs in the event; Ns-- stripes number of ZPs in the event; Ref-- reference; S-- the symbols used in Figure \ref{fig-2}.}
\tablerefs{(1) Huang et al. 2008; (2) Chernov et al. 2003; (3) Chernov et al. 2005; (4) Yan et al. 2007}
\end{deluxetable}

\section{OBSERVATIONS AND DATA ANALYSIS}\label{sec-2}
\begin{figure}[htbp!]
\epsscale{0.6}
\plotone{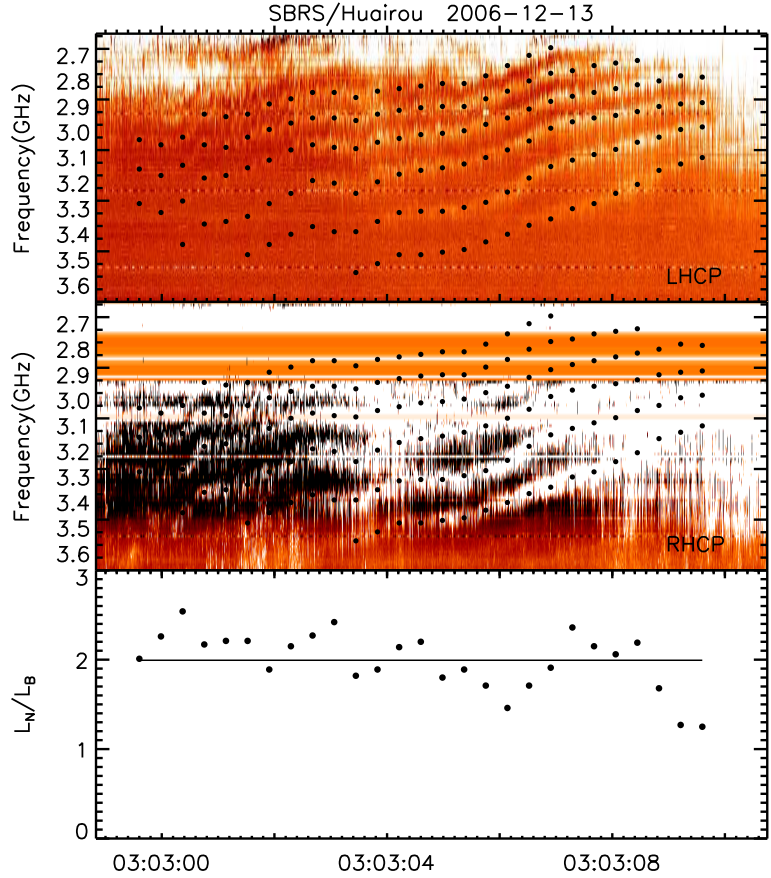}
\caption{ZP occurred at 2.6--3.8 GHz bandwidth recorded by SBRS/Huairou in 03:03:00--03:03:10 UT during the X3.4 flare on 2006 December 13. The \emph{upper and middle panels} are for the LHCP and RHCP components, respectively. The frequencies $f$ of the ZP stripes at every sampling-time are marked with filled circle dots. The \emph{lower panel} demonstrates the values of $L_N/L_B$ at every sampling-time with filled circle dots. The horizontal line denotes the average value of $L_N/L_B$ in the whole ZP structure, taken as the ratio $L_N/L_B$ of this ZP.
(A color version of this figure is available in the online journal.) \label{fig-0}}
\end{figure}

Figure \ref{fig-0} shows the microwave dynamic spectrograms of a typical ZP structure on left- and right-handed circular polarization (LHCP and RHCP) recorded on 13 December 2006. The microwave spectrograms required for this study are obtained by SBRS/Huairou. SBRS/Huairou is an advanced solar radio spectrometer that measures the total flux density of solar radio emission with dual circular polarization, i.e., the LHCP and RHCP. It consists of three parts: 1.10--2.06 GHz (cadence of 5 ms and frequency resolution of 4 MHz), 2.60--3.80 GHz (cadence of 8 ms and frequency resolution of 10 MHz), and 5.20--7.60 GHz (cadence of 5 ms and frequency resolution of 20 MHz). The observation sensitivity is $S/S_{\sun}\leqslant 2\%$, where $S_{\sun}$ is quiet solar background radiation \citep{2002ESASP.506..375Y,2004SoPh..222..167F}. During the solar cycle 23, SBRS/Huairou had obtained excellent observation data at 2.6--3.8 GHz bandwidth, and numerous ZPs are observed at this bandwidth. We found 74 ZPs in 9 solar flares, whose $L_N/L_B$ can be calculated, observed by SBRS/Huairou at 2.6--3.8 GHz from 1997 to 2006. The parameters of the 9 flare events are summarized in Table \ref{tbl-1}. Among the listed 9 events, we found 3 typical flares (2006 December 13, 2002 April 21, and 2001 October 19) that legible ZPs appeared in both of before and after the soft X-ray maximums in flaring processes. In the others 6 events, ZPs appeared either prior or after the flaring peaks. It should be noted that, in the eleventh column in Table \ref{tbl-1},  the number of stripes of each ZPs is mainly below 10, while in the event of 2002 April 21, the number is up to 34.

We now consider that the DPR model is responsible for the emitting of ZP. It is reasonable to assume both the magnetic field strength ($B$) and the plasma density ($n_e$) decrease exponentially with height ($h$) along the flux tube in the corona:
\begin{equation}\label{equ-1}
B=B_0e^{-\Delta h/L_B},n_e=n_{e_0}e^{-\Delta h/L_N} ,
\end{equation}
where $B_0$ and $n_{e_0}$ are the magnetic field and the plasma density at $h_0$ respectively, and $L_N = n_e (\partial n_e/\partial h)^{-1}$, $L_B = B (\partial B/\partial h)^{-1}$. The electron cyclotron frequency $f_{ce}$ and the plasma frequency of electrons $f_{pe}$ can be expressed as:
\begin{equation}\label{equ-2}
f_{ce}=f_{ce_0}e^{-\Delta h/L_B}, f_{pe}=f_{pe_0}e^{-\Delta h/(2L_N)}
\end{equation}

According to DPR model \citep{2007SoPh..241..127K}, the frequency separation $\Delta f_s=f_{s+1}-f_s$, between the adjacent ZP stripes at harmonics $s$ and $s+1$, is related to $B$ and $n_e$,
\begin{equation}\label{equ-3}
\frac{\Delta f_s}{f_s} = \approx \frac{f_{pe}(h_{s+1}) - f_{pe}(h_s)}{f_{pe}(h_s)} \approx \frac{1}{s} \frac{1}{1-(2L_N/L_B)}
\end{equation}
where $h_{s+1}$ and $h_s$ are the heights of neighboring resonance levels $s+1$ and $s$. For a given ZP, with zebra stripes corresponding to harmonics $s_1$ and $s_2$, we can determine the frequencies of the two stripes at given time. Denoting $\Delta s = s_2 - s_1$ and $\delta_s = \Delta f_s/f_s$, Equation (\ref{equ-3}) can be transformed to
\begin{equation}\label{equ-4}
  s_1 = \frac{\Delta s\, \delta _{s_2}}{\delta _{s_1} - \delta _{s_2}} ,\qquad
  \frac{L_N}{L_B} \approx \frac{1}{2} \left(1 + \frac{1}{|\delta _{s_1}| s_1}\right) .
\end{equation}
Therefore the ratio of $L_N/L_B$ can be calculated from the observation of ZP by the DPR model. Note that the $\delta_s$ in Equation (\ref{equ-4}), the relative frequency separation between two neighboring stripes, restricts that a ZP should possess at least 3 legible stripes in order to calculate the value of $L_N/L_B$.

\begin{figure}[htbp!]
\epsscale{0.6}
\plotone{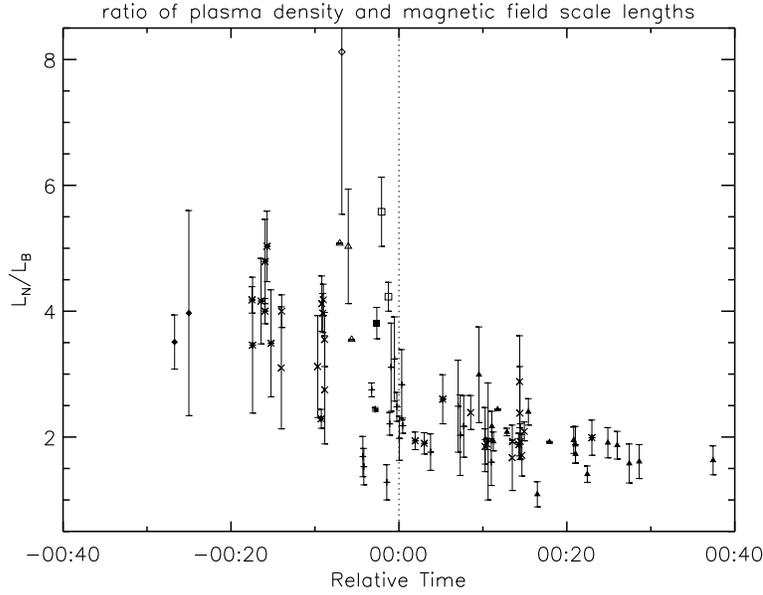}
\caption{$L_N/L_B$ estimated from 74 microwave ZPs in the 9 flare events listed in Table \ref{tbl-1}, are plotted against the time relative to the GOES soft X-ray flaring peaks (\emph{vertical dotted line}) of each flares. The symbols used to denote the ratio $L_N/L_B$ are also listed in Table \ref{tbl-1}.  \label{fig-2}}
\end{figure}

Figure \ref{fig-0} illustrates the method we employed to obtain the $L_N/L_B$ of a typical ZP. This ZP was recorded at 03:03:00 UT on 13 December 2006. It lasted for about 7 minutes, but only the part during 03:03:00--03:03:10 UT can be well discriminated from the spectrograms due to saturation in the frequency range of 2.75--2.90 GHz. The upper two panels are the dynamic spectrograms of the ZP on LHCP and RHCP. Six stripes can be well discriminated from the background emission in this case. We applied a sufficient sampling rate to calculate the $L_N/L_B$ of every ZP, depending on the time duration of the ZP. For a given time, the frequency $f$ of one stripe is determined as the center-frequency of the stripe at the sampling-time, denoted with filled circle dots. Equation (\ref{equ-4}) implies that the larger the difference between the harmonics of two stripes ($\Delta s$) is, the smaller the $L_N/L_B$ error caused by frequency determination is. Therefore, we use the lowermost two and the uppermost two legible stripes to calculate $L_N/L_B$. The \emph{lower panel} demonstrates the calculated values of $L_N/L_B$ at all sampling-time with filled circle dots. The horizontal line denotes the average value of $L_N/L_B$ in the whole ZP structure, which is 1.99 in this case, taken as the typical ratio $L_N/L_B$ of this ZP.

\section{RESULTS}\label{sec-3}
In Figure \ref{fig-2}, we plot the ratio $L_N/L_B$ derived from 74 ZPs listed in Table \ref{tbl-1}, against the time relative to the GOES soft X-ray flaring peaks of each flares. The symbols used to denote the ratio in different events are shown in the last column of Table \ref{tbl-1}. The standard deviation $\sigma$ of ZPs are shown as error bars. The irregularly drift of ZP stripes indicates that the resonance levels do not remain at precisely the same height in flux tube, which may account for the $\sigma$. The time of the GOES soft X-ray peaks are assigned as 00:00, marked with the \emph{vertical dotted line}. The temporal variation of the ratio (Fig.\ref{fig-2}) shows that (1) during the flaring processes of these flares, the ratio is mainly in the range of between 1.5 and 5; (2) before the flaring peak, the ratio is mainly in range of 3--5, except for the event of 2002 April 21 (\emph{pluses}); after the peak, the ratio decreases to 1--3; (3) in the events of 2006 December 13 (\emph{asterisk}), 2002 April 21, and 2001 October 19 (\emph{times}), whose ZPs appeared in both of before and after the flaring peaks, the ratio shows interesting temporal variation during the flaring process.

\begin{figure}[htbp!]
\epsscale{0.82}
\plotone{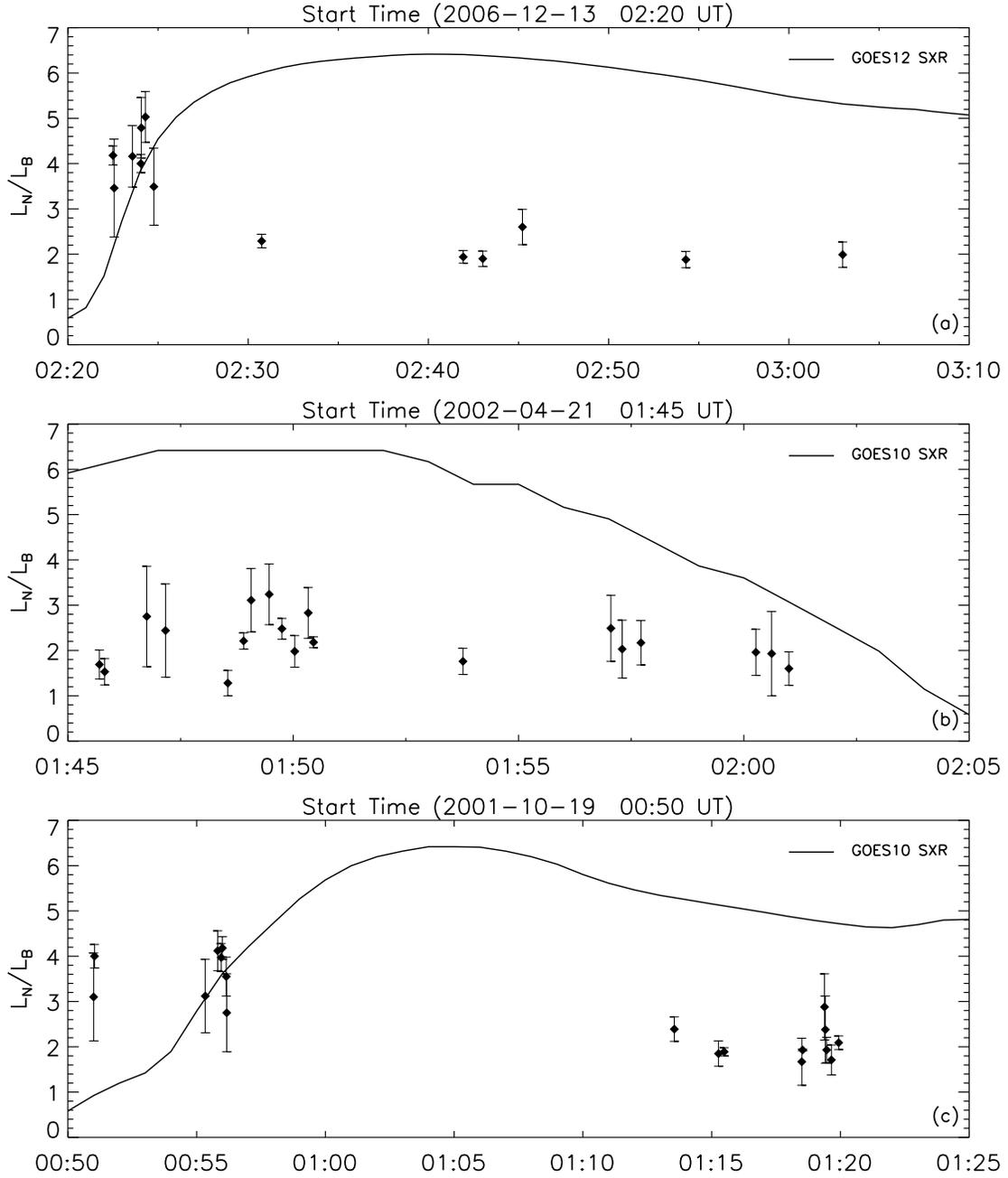}
\caption{Temporal variation of the ratio $L_N/L_B$ in the flaring process of the three flare events: (a) 2006 December 13; (b) 2002 April 21; (c) 2001 October 19. The \emph{soild curves} denote the normalized GOES soft X-ray profiles at 1-8 \AA. The \emph{vertical dotted line} refers to the time of the X-ray peak. The \emph{pluses} denote the values of $L_N/L_B$ at each sampling-time and the ratio $L_N/L_B$ of each ZP case is given by \emph{diamonds}. The error bar denotes the $\sigma$ of $L_N/L_B$. \label{fig-3}}
\end{figure}

Figure \ref{fig-3} compares the temporal variation of the ratio $L_N/L_B$ in the flaring process of the three flare events. The ratio of each ZP case is given by \emph{filled diamond}. The \emph{soild curve} denote the normalized GOES soft X-ray profile at 1-8 \AA. The \emph{vertical dotted line} refers to the time of the soft X-ray peak. The error bar denotes the $\sigma$ of $L_N/L_B$. The first impression of Figure \ref{fig-3} is that the ratio shows decrease trend during the flaring process of each event. It is also noted that the ratio displays a step-wise decrease around the soft X-ray peak in the events of 2006 December 13 and 2001 October 19, while the ratio decreases slowly during the flaring process in the event of 2002 April 21. We will analyse the 3 typical flare events in detail separately in following text.

\begin{figure}[htbp!]
\epsscale{0.82}
\plotone{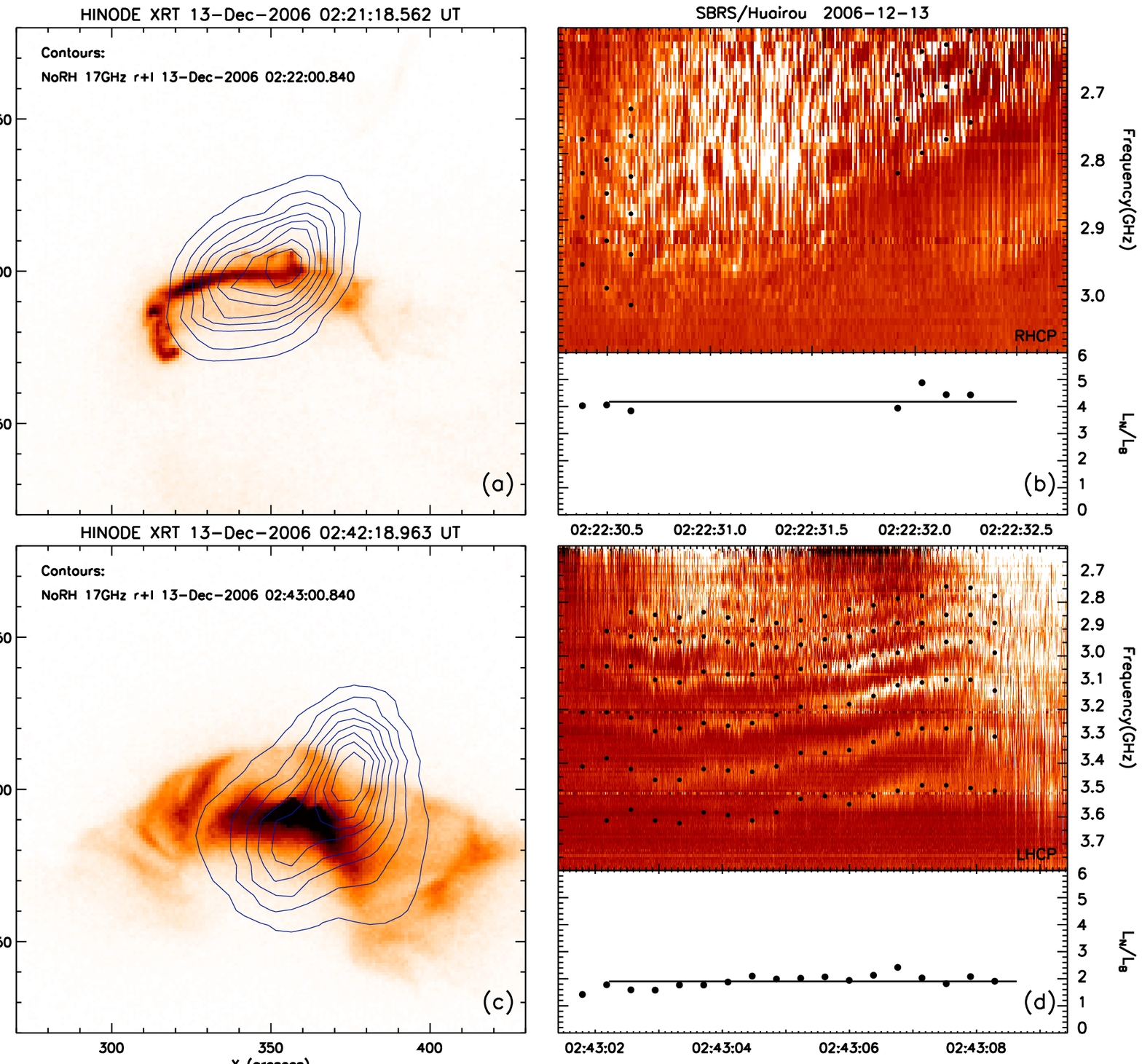}
\caption{NoRH 17 GHz contours superposed on the Hinode/XRT images of NOAA 10930 at 13 December 2006, taken before (02:21 UT; a) and after (02:42 UT; c) the step-wise decrease of ratio $L_N/L_B$. The corresponding ZPs and their $L_N/L_B$ are shown in panel b and d.
(A color version of this figure is available in the online journal.) \label{fig-4}}
\end{figure}

\subsection{The event of 2006 December 13}\label{sec-3.1}
On 2006 December 13, an X3.4/4B class, typical two-ribbon flare occurred \citep{2007PASJ...59S.807I} in the active region NOAA 10930 (S05W33). \citet{2007PASJ...59S.785S} presents the long term evolution of the sheared magnetic fields in this active region by multi-wavelength observations. The flare started at 02:14 UT, reached maximum at around 02:40 UT, and ended at 02:57 UT. Here, we are interested in the period of 02:22--03:05 UT because all 13 ZPs of this flare event were recorded during this period. A clear step-wise decrease before the flaring peak divides the process into two phases: in the first phase (before the step-wise decrease), 7 ZPs are observed and the ratio $L_N/L_B$ is 3.50--5.00; the other 6 ZPs are observed in the second phase (after the step-wise decrease) and the ratio decreases to 1.85--2.65. In Figure 4, the left column compares the X-ray coronal structure in the flaring region taken by the X-ray Telescope (XRT) aboard Hinode before (a) and after (c) the step-wise decrease. The contours show the microwave intensity at 17 GHz observed by Nobeyama Radio Heliograph (NoRH). The first phase (Fig.\ref{fig-4}a) is characterized by a east-west, highly sheared, bundle of nearly parallel loops. The ratio fluctuates in the range between 3.50 and 5.00 (Fig.\ref{fig-3}a). Figure \ref{fig-4}b shows a ZP started at 02:22:30 UT and the ratio is 3.90. Similar to Figure \ref{fig-2}, but only one spectrogram on RHCP is demonstrated, based on the polarization of the ZP. At the end of the first phase, significant energy release is processing as several saturated brightenings showing up along the elongated structure between 02:22:18--02:24:18 UT. Also noted is the coronal structure underwent a topological change that a north-south, less sheared, expanding arcade structure (Fig.\ref{fig-4}c) was formed. In the second phase, the ratio has decreased to around 2.00 (Fig.\ref{fig-4}d). The contours of NoRH 17 GHz intensity (Fig.\ref{fig-4}a and \ref{fig-4}c) show that the microwave sources in the two phases have approximately the same positions, coincident with the position of the X-ray loops structure, supposes that the radiation of ZPs may come from the X-ray bright loops.

\begin{figure}[htbp!]
\epsscale{0.82}
\plotone{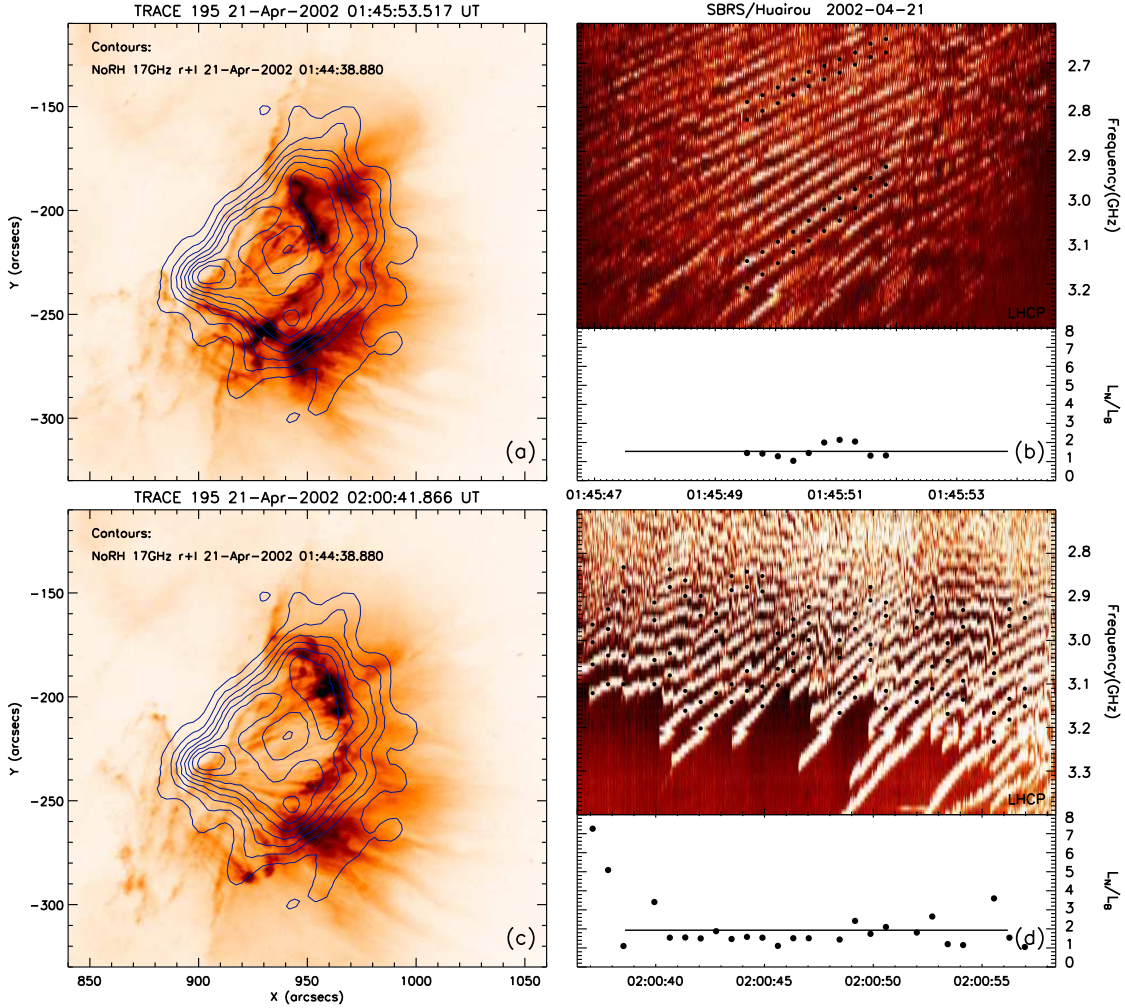}
\caption{NoRH 17 GHz contours superposed on the TRACE/195 \AA\ images of NOAA 9906 at 21 April 2002, taken before (01:45 UT; a) and after (02:00 UT; c) flaring peak. The corresponding ZPs and their $L_N/L_B$ are shown in panel b and d.
(A color version of this figure is available in the online journal.)  \label{fig-5}}
\end{figure}

\subsection{The event of 2002 April 21}\label{sec-3.2}

On 2002 April 21, an X1.5/1F class, two ribbons flare occurred in active region NOAA 9906 close at the western solar limb (S14W84). The GOES soft X-ray emission shows that the flare started at 00:43 UT, reached to its maximum at around 01:50 UT, and ended at 02:38 UT. A detail analysis of the X1.5 flare was given by \citet{2002SoPh..210..341G}. ZPs in this event (Fig.\ref{fig-5}b and \ref{fig-5}d) possess intricate and long-lasting stripes \citep{2004IAUS..219..722Y}, which are far more complicated than the ZPs in the 2006 December 13 event and the 2001 October 19 event. About 10 ZPs were observed in the time interval of 01:40--02:05 UT. It is worth mentioning that, for instance, the third and forth "ZPs" in this event are not two separate ZPs, but one long-lasting ZP structure with fast-varying frequency parameter, therefore we treated the ZP as two separate ZPs.

Figure \ref{fig-3}b shows the variation of the ratio $L_N/L_B$ in this flare event. Different from the "two-phase" pattern in the 2006 December 13 event, the ratio decreases slowly during the flaring process. Figure \ref{fig-5} illustrates the Transition Region and Coronal Explorer (TRACE) EUV images at 195 \AA\ superposed with NoRH 17 GHz contours, and the corresponding ZPs in this event. A rising post-flare loop system, at a rate of about 10 km s$^{-1}$ \citep{2002SoPh..210..341G}, was seen during 01:40--02:05 UT. By comparing the coronal structure\, we found that the coronal configuration does not show so significant change in the flaring process. The contours of NoRH 17 GHz show the emission mainly came from the foot-point and the top of the coronal loops. The Siberian Solar Radio Telescope (SSRT) data at 5.7 GHz indicates that microwave source was located on the top of the coronal loops \citep{2005A&A...437.1047C}.

\begin{figure}[htbp!]
\epsscale{0.82}
\plotone{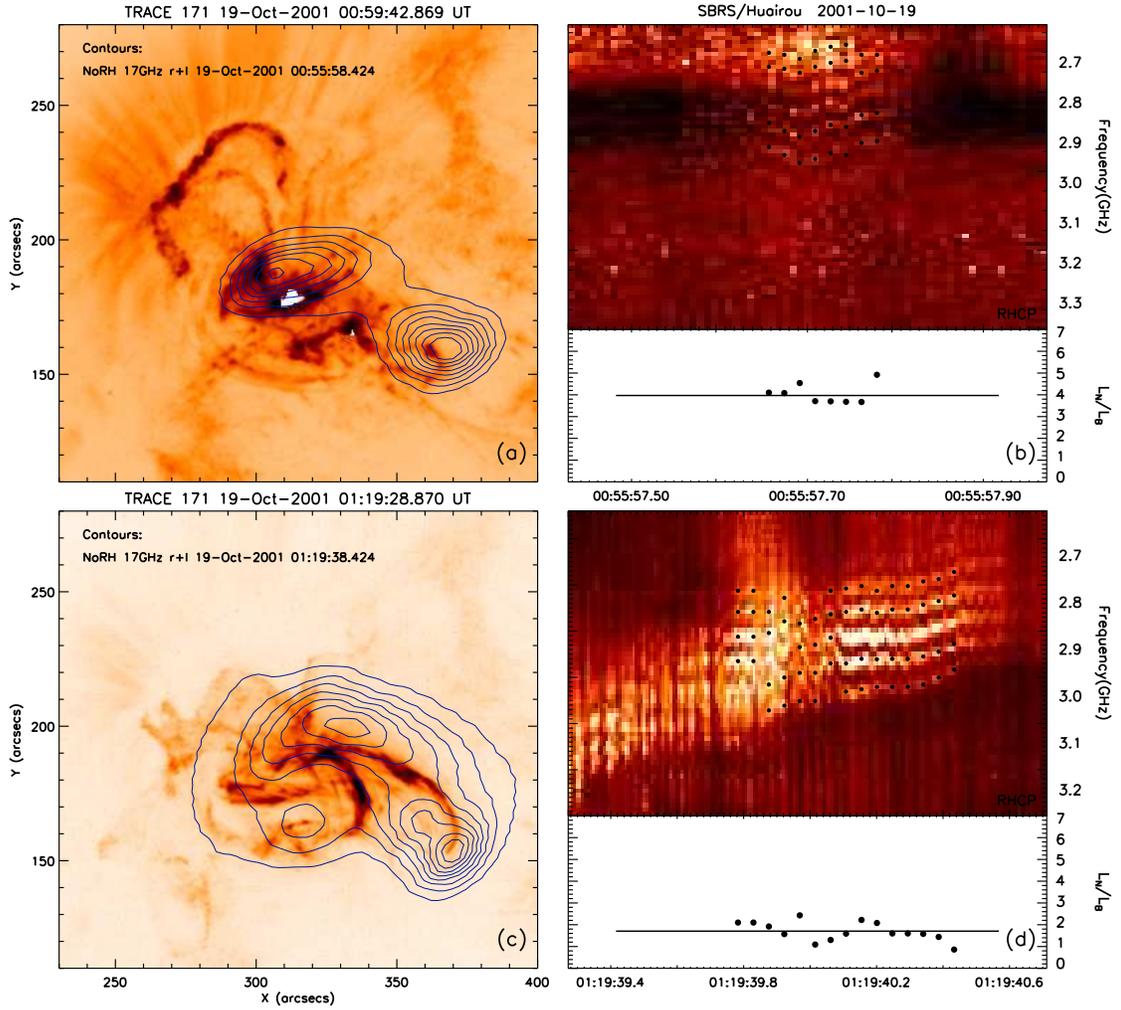}
\caption{NoRH 17 GHz contours superposed on the TRACE/171 \AA\ images of NOAA 9661 at 19 October 2001, taken before (00:59 UT; a) and after (01:19 UT; c) the step-wise decrease of ratio $L_N/L_B$. The corresponding ZPs and their $L_N/L_B$ are shown in panel b and d.
(A color version of this figure is available in the online journal.) \label{fig-6}}
\end{figure}

\subsection{The event of 2001 October 19}\label{sec-3.3}
A X1.6/2B class two-ribbon flare occurred in active region NOAA 9661 (N16W18) around 00:47 UT on 2001 October 19. A detailed study on multi-wavelength involved in this event was carried out by \citet{2004ApJ...606..583L}. The flare started at 00:47 UT, reached maximum at around 01:05 UT, and ended at 01:13 UT. 18 ZPs were recorded in the time interval 00:50--01:20 UT.

The variation of the ratio $L_N/L_B$ in this event (Fig.\ref{fig-3}c) shows a similar "two-phase" pattern occurred in the event of 13 December 2006. In the first phase, 8 ZPs were recorded and the ratio is 2.75--4.19; the other 10 ZPs are observed in the second phase and the ratio decreases to 1.67--3.17. Figure \ref{fig-6} illustrate the variation of the configuration of coronal loops during this flare and the corresponding ZPs. Here, we present the TRACE 171 \AA\ images to show the configuration of the coronal loops, which can be represented the variation of the coronal magnetic topology. Two ribbons structure (Fig.\ref{fig-6}a) is observed in the first phase, during which the ratio concentrates in the range of 3.00--4.00. The TRACE 171 \AA\ images imply that the magnetic field topology changed rapidly at 01:09:59 UT. No ZP was recorded in the transition between the first and the second phase. In the second phase, the coronal loops evolved to a expanding post-flare loops configuration in 171 \AA\ images (Fig.\ref{fig-6}c). Figure \ref{fig-6}d shows the corresponding ZP appeared at 01:19 UT and the ratio is around 2.00. The contours of NoRH 17 GHz overlaying the TRACE 171 \AA\ images show a loop structure stretching across the two flare ribbons. At around 00:55 UT, the microwave emission originated mainly from the two foot-points of the coronal loops. About 3 minutes later, the microwave sources gradually converged to the loop top and changed little morphologically thereafter.

\section{DISCUSSIONS AND CONSLUSION}\label{sec-4}
\begin{figure}[htbp!]
\epsscale{0.5}
\plotone{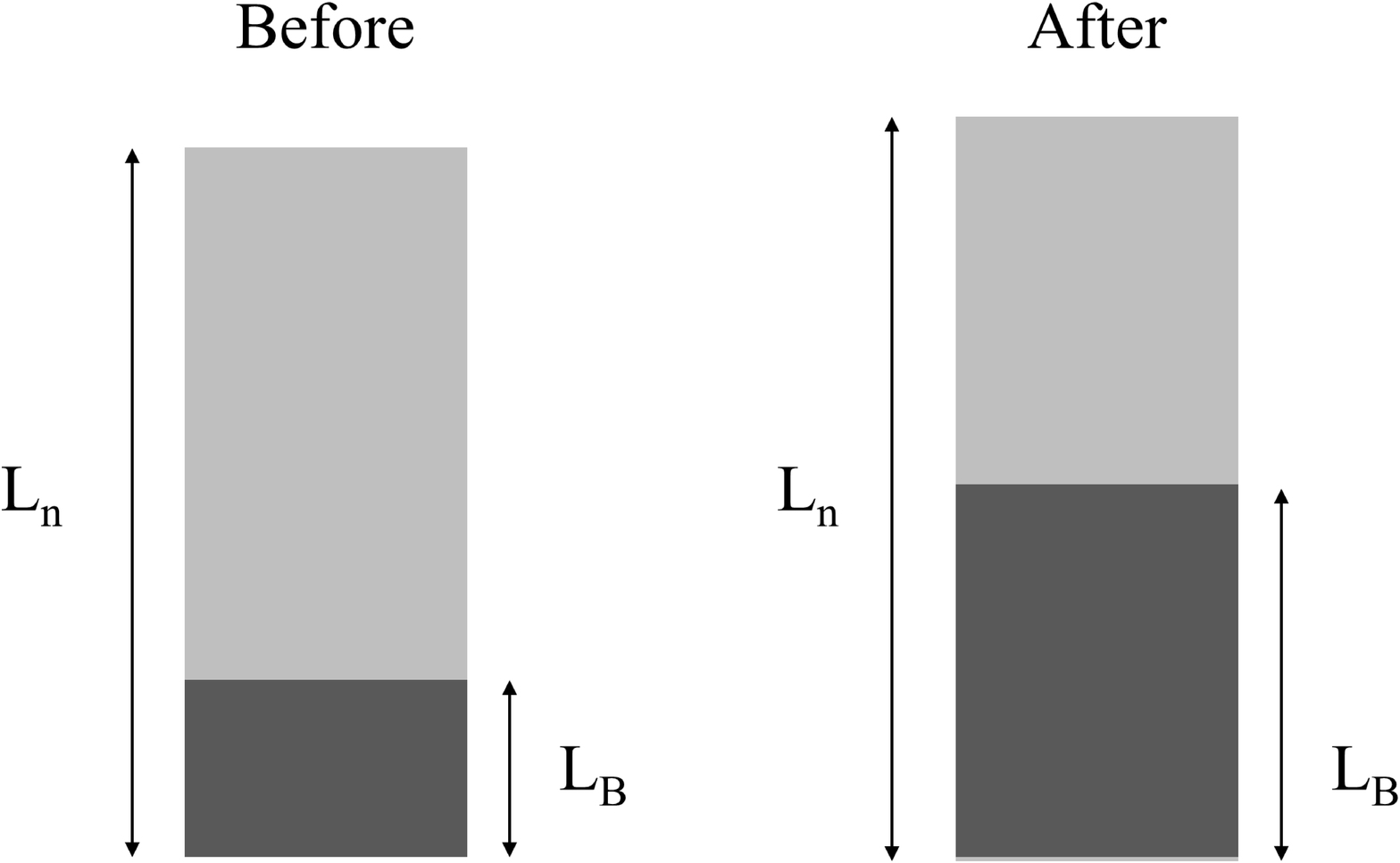}
\caption{Illustration of the ratio $L_N/L_B$ before and after the topological change during the flaring process.  \label{fig-7}}
\end{figure}

In this paper, We investigated the variation of 74 microwave ZPs recorded by SBRS/Huairou at 2.6--3.8 GHz in 9 eruptive solar flares. By analyzing ZP stripes appeared in flares, we calculated the ratio $L_N/L_B$ in emission sources. We have obtained following results.

For all the cases, the ratio $L_N/L_B$ prior to the flaring peaks tend to be larger than those after the peaks. The ratio mainly concentrate in the range of 3-5 before the flaring peaks except for the case of 2002 April 21. After the free energy releases, the ratio drops to about 2. Among these cases, we found 3 typical flares whose ZPs appeared in both of before and after the flaring peaks. In the cases of 2006 December 13 and 2001 October 19, the ratio displays a step-wise decrease during the processes of energy releasing. In the case of 2002 April 21, the ratio shows a tendency of slow decrease. The decrease of the ratio reveals that the magnetic field scale height increases faster relative to the plasma density scale height during the flaring process, implies the presence of a relaxation process of the magnetic field relative to the plasma density.

We compare the X-ray/EUV images taken before and after the step-wise decrease of the ratio $L_N/L_B$ of the three events. The coronal configuration underwent a topological change after the release of the stored free energy, reveals a sigmoid-to-arcade transformation \citep{2005A&A...434..725M,2007ApJ...658L.127L,2007ApJ...664L.127J} during flaring process in the events of 2006 December 13 and 2001 October 19. As well accepted, the shape of X-ray/EUV loops infers the current-free magnetic field lines. The loop arcades after the topological change showing a relaxed form of magnetic field configuration, implies that the topological change of magnetic field in flaring region is a process toward a relaxed, potential state by releasing the magnetic shear \citep{1992PASJ...44L.123S}. In the case of 2002 April 21, the EUV images do not see so significant change in coronal magnetic field configuration in this event. We check the NoRH observations of these three flare events. The microwave sources of ZPs were located on the bright flare loops observed at EUV/X-ray band. The height of source regions of ZPs were estimated to be about 66 Mm \citep{2012ApJ...745..186H}, right in the altitude range of $\sim\!8$ Mm to $\sim\!70$ Mm where the magnetic nonpotentiality decreases after the flare event of 2006 December 13 \citep{2008ApJ...676L..81J}. This entirely agreement supports the validity of the relation between the $L_N/L_B$ variation and the energy release. When we suppose that the $L_N/L_B$ is determined by the geometrical structure of magnetic field and the plasma density, then the step-wise decreases of $L_N/L_B$ (Fig.\ref{fig-3}a and \ref{fig-3}c) may be related to the topological change on the coronal magnetic field structure, whereas the slowly varying $L_N/L_B$ (Fig.\ref{fig-3}b) infers no significant change of coronal magnetic field, then our results favor the $L_N/L_B$ variation being due to the topological change of flare core region where the main part of the magnetic energy is released. Figure \ref{fig-7} illustrates the picture of the ratio before and after the topological change. For a typical value of the ratio, the magnetic field scale height $L_B$ is 1/4 of the plasma density scale height $L_N$ before the topological change. After the topological change processing, $L_B$ becomes 1/2 of $L_N$. This result may constrain the solar flare modeling to some extent.

The ratio $L_N/L_B$ plays a significant role in the DPR model \citep{1975SoPh...44..461Z}. However the absolute values of $L_N$ and $L_B$ are even more important. The smaller these values are, the more stripes of a ZP can be realized in certain frequency range.  We noted that, from Table \ref{tbl-1}, the numbers of zebra stripes varied from event to event, especially for the event of 2002 April 21 which is up to 34. In other words, the absolute values of $L_N$ and $L_B$ varied in different events.  Whereas, the ratio $L_N/L_B$ is mainly in the range of 1.5--5, as shown in Figure \ref{fig-2}, which is independent from the varied absolute values of $L_N$ and $L_B$. 

Here, it is worth discussing the validity in particular flare event, e.g., the event of 2002 April 21 with the fact that up to 34 ZP stripes existed simultaneously in a short frequency range. In this event, for $\Delta f_s  \sim\!40$ MHz,  we could obtain the magnetic field  $B = 19.5$ Gs, calculated from equation (\ref{equ-3}) $\frac{\Delta f_s}{f_{ce}}  \approx \frac{1}{1-(2L_N/L_B)}$, which is corresponding to altitude of $\sim\!60$ Mm \citep{1978SoPh...57..279D}. While from the plasma frequency $f_{pe} \approx 8.98 \times 10^3 \sqrt n_e$ and $\beta = nk_BT/(\frac{B^2}{8\pi})$, where $k_B$ is the Boltzmann's constant and $n\sim n_e$, one can get $\beta \approx 1.4$ for $T=1.5\times 10^6 K$. The value of $\beta$ seems over high at this altitude of $60$ Mm, according to current knowledge.  Thus, this is a major obstacle for the DPR model to produce 34 ZP stripes simultaneously at 2.6--3.8 GHz range in this event.  Similar calculation had been carried out showing that the DPR model fails to interpret the formation of a large number of ZP stripes \citep{2010RAA....10..821C}.  WW model was proposed to be responsible for the formation of the multi-stripes ZP structures in this event, as the ratio $L_N/L_B$ is not a significant parameter in the WW model, but the absolute value of magnetic field B is. The magnetic field  B can be estimated to be a reasonable value of $71.5$ Gs with the formula $\Delta f=0.2f_{ce}$, where $\Delta f$ is the frequency separation between two adjacent ZP stripes \citep{2006SSRv..127..195C}. 

Finally, it needs to be mentioned that it deserves further investigation as to answer why the $L_N/L_B$ is 3--5 and 2, before and after the energy releasing process, respectively. Additionally, although the radio heliograph observations suggest that the microwave ZPs may come from the flare loops, we have no enough evidence to confirm that the ZPs originated from the same position in flare loops because it is difficult to obtain the accurate position of radio source region due to a lack of high-spatial and high-temporal radio imaging observation in the corresponding frequency range (2.6--3.8 GHz). The constructing high-spatial and high-temporal resolution Chinese Spectral Radio Heliograph \citep[CSRH, 0.4--15 GHz;][]{2009EM&P..104...97Y} will provide an opportunity to give some clear answers to the above problems.

\acknowledgments

The authors would like to thank to G. P. Chernov for suggestive comments.  B. C. Low is acknowledged for helpful discussion. The authors would also like to thank the Hinode/XRT, NoRH and TRACE teams for providing the excellent data. This work is supported by NFSC Grant No.10921303, 11273030, MOST Grant No.2011CB811401, and the National Major Scientific Equipment R\&D Project ZDYZ2009-3.


\begin{thebibliography}{}

\bibitem[Altyntsev et
al.(2005)]{2005A&A...431.1037A} Altyntsev, A.~T., Kuznetsov, A.~A., Meshalkina, N.~S., Rudenko, G.~V., \& Yan, Y.~H.\ 2005, \aap, 431, 1037



\bibitem[Bastian(2004)]{2004ASSL..314...47B} Bastian, T.~S.\ 2004,
Astrophysics and Space Science Library, 314, 47


\bibitem[Bastian et
al.(1998)]{1998ARA&A..36..131B} Bastian, T.~S., Benz, A.~O., \& Gary, D.~E.\ 1998, \araa, 36, 131


\bibitem[Chernov et
al.(2003)]{2003A&A...406.1071C} Chernov, G.~P., Yan, Y.~H., \& Fu, Q.~J.\ 2003, \aap, 406, 1071


\bibitem[Chernov et
al.(2005)]{2005A&A...437.1047C} Chernov, G.~P., Yan, Y.~H., Fu, Q.~J., \& Tan, C.~M.\ 2005, \aap, 437, 1047


\bibitem[Chernov(2006)]{2006SSRv..127..195C} Chernov, G.~P.\ 2006, \ssr,
127, 195


\bibitem[Chernov(2010)]{2010RAA....10..821C} Chernov, G.~P.\ 2010, Research
in Astronomy and Astrophysics, 10, 821


\bibitem[Chiuderi et al.(1973)]{1973SoPh...33..225C} Chiuderi, C.,
Giachetti, R., \& Rosenberg, H.\ 1973, \solphys, 33, 225


\bibitem[Dulk 
\& McLean(1978)]{1978SoPh...57..279D} Dulk, G.~A., \& McLean, D.~J.\ 1978, \solphys, 57, 279 


\bibitem[Elgar{\"o}y(1959)]{1959Natur.184..887E} Elgar{\"o}y, O.\ 1959,
\nat, 184, 887


\bibitem[Fu et al.(2004)]{2004SoPh..222..167F} Fu, Q.~J., Ji, H.~Q., Qin, Z.~H., et
al.\ 2004, \solphys, 222, 167


\bibitem[Gallagher et al.(2002)]{2002SoPh..210..341G} Gallagher, P.~T.,
Dennis, B.~R., Krucker, S., Schwartz, R.~A.,
\& Tolbert, A.~K.\ 2002, \solphys, 210, 341


\bibitem[Huang
\& Tan(2012)]{2012ApJ...745..186H} Huang, J., \& Tan, B.~L.\ 2012, \apj, 745, 186


\bibitem[Hudson(2011)]{2011SSRv..158....5H} Hudson, H.~S.\ 2011, \ssr, 158,
5


\bibitem[Isobe et al.(2007)]{2007PASJ...59S.807I} Isobe, H., Kubo, M.,
Minoshima, T., et al.\ 2007, \pasj, 59, 807


\bibitem[Jing et al.(2007)]{2007ApJ...664L.127J} Jing, J., Lee, J., Liu,
C., Gary, D.~E., \& Wang, H.~M.\ 2007, \apjl, 664, L127


\bibitem[Jing et al.(2008)]{2008ApJ...676L..81J} Jing, J., Wiegelmann, T.,
Suematsu, Y., Kubo, M., \& Wang, H.~M.\ 2008, \apjl, 676, L81


\bibitem[Kuijpers(1975)]{1975A&A....40..405K} Kuijpers, J.\ 1975, \aap, 40, 405


\bibitem[Kuznetsov
\& Tsap(2007)]{2007SoPh..241..127K} Kuznetsov, A.~A., \& Tsap, Y.~T.\ 2007, \solphys, 241, 127


\bibitem[Li
\& Ding(2004)]{2004ApJ...606..583L} Li, J.~P., \& Ding, M.~D.\ 2004, \apj, 606, 583


\bibitem[Liu et al.(2007)]{2007ApJ...658L.127L} Liu, C., Lee, J., Gary,
D.~E., \& Wang, H.\ 2007, \apjl, 658, L127


\bibitem[Longcope(2005)]{2005LRSP....2....7L} Longcope, D.~W.\ 2005, Living
Reviews in Solar Physics, 2, 7


\bibitem[Low(2006)]{2006ApJ...649.1064L} Low, B.~C.\ 2006, \apj, 649, 1064


\bibitem[Mandrini et
al.(2005)]{2005A&A...434..725M} Mandrini, C.~H., Pohjolainen, S., Dasso, S., et al.\ 2005, \aap, 434, 725


\bibitem[Ning et
al.(2000)]{2000A&A...364..853N} Ning, Z.~J., Fu, Q.~J., \& Lu, Q.\ 2000, \aap, 364, 853


\bibitem[Priest
\& Forbes(2002)]{2002A&ARv..10..313P} Priest, E.~R., \& Forbes, T.~G.\ 2002, \aapr, 10, 313


\bibitem[Rosenberg(1972)]{1972SoPh...25..188R} Rosenberg, H.\ 1972,
\solphys, 25, 188


\bibitem[Sakurai et al.(1992)]{1992PASJ...44L.123S} Sakurai, T., Shibata,
K., Ichimoto, K., Tsuneta, S., \& Acton, L.~W.\ 1992, \pasj, 44, L123


\bibitem[Shibata
\& Magara(2011)]{2011LRSP....8....6S} Shibata, K., \& Magara, T.\ 2011, Living Reviews in Solar Physics, 8, 6


\bibitem[Slottje(1972)]{1972SoPh...25..210S} Slottje, C.\ 1972, \solphys,
25, 210


\bibitem[Su et al.(2007)]{2007PASJ...59S.785S} Su, Y.~N., Golub, L., van
Ballegooijen, A., et al.\ 2007, \pasj, 59, 785


\bibitem[Tan et al.(2012)]{2012ApJ...744..166T} Tan, B.~L., Yan, Y.~H., Tan, C.~M.,
Sych, R., \& Gao, G.\ 2012, \apj, 744, 166


\bibitem[Yan et al.(2002)]{2002ESASP.506..375Y} Yan, Y.~H., Fu, Q.~J., Liu, Y.~Y.,
\& Chen, Z.~J.\ 2002, ESASP, 506, 375


\bibitem[Yan et
al.(2009)]{2009EM&P..104...97Y} Yan, Y.~H., Zhang, J., Wang, W., Liu, F., Chen, Z., Ji, G., et al.\ 2009, Earth Moon and Planets, 104, 97


\bibitem[Yan et al.(2007)]{2007PASJ...59S.815Y} Yan, Y.~H., Huang, J., Chen,
B., \& Sakurai, T.\ 2007, \pasj, 59, 815


\bibitem[Yan et al.(2004)]{2004IAUS..219..722Y} Yan, Y.~H., Zhang, L., Tan,
C.~M., et al.\ 2004, IAUS, 219, 722


\bibitem[Zheleznyakov
\& Zlotnik(1975)]{1975SoPh...44..461Z} Zheleznyakov, V.~V., \& Zlotnik, E.~Y.\ 1975, \solphys, 44, 461


\bibitem[Zlotnik(1977)]{1977SvA....21..744Z} Zlotnik, E.~I.\ 1977, \sovast,
21, 744



\end{thebibliography}
\end{document}